\documentclass[fleqn,usenatbib]{mnras}

\usepackage[english]{babel}
\usepackage{amsmath}
\usepackage{amssymb}
\usepackage{graphicx}
\usepackage{subfigure}
\usepackage[normalem]{ulem}

\usepackage{verbatim}

\usepackage{color}
\usepackage{multirow}
\usepackage{mathtools}
\usepackage{epstopdf}

\usepackage{url}
\usepackage{xcolor}

\def\mnras{MNRAS}
\def\apj{ApJ}

\def\aap{A\&A}
\def\aaps{A\&A Supp.}

\def\araa{ARA\&A}

\graphicspath{{figures/}{../figures/}}

\usepackage{newtxtext,newtxmath}
\usepackage[T1]{fontenc}

\DeclareRobustCommand{\VAN}[3]{#2}
\let\VANthebibliography\thebibliography
\def\thebibliography{\DeclareRobustCommand{\VAN}[3]{##3}\VANthebibliography}

\usepackage{graphicx}	% Including figure files
\usepackage{amsmath}	% Advanced maths commands
\usepackage{amssymb}	% Extra maths symbols

\usepackage{hyperref} % link in bibliografia
\usepackage{natbib}

\newcommand{\msun}{\ensuremath{\rm{M_{\odot}}}}

\title[]{Understanding the origin of CEMP-no stars through ultra-faint dwarfs}

\author[M. Rossi et al.]{
Martina Rossi$^{1,2}$\thanks{E-mail:martina.rossi@unifi.it},
Stefania Salvadori$^{1,2}$,
{\'A}sa Sk{\'u}lad{\'o}ttir$^{1,2}$,
Irene Vanni$^{1,2}$
\\
$^{1}$Dipartimento di Fisica e Astrofisica, Universitá degli Studi di Firenze, via G. Sansone 1,Sesto Fiorentino,Italy\\
$^{2}$INAF/Osservatorio Astrofisico di Arcetri, Largo E. Fermi 5, Firenze, Italy
}

\date{Accepted February 2023 , Received 2023 , in original form December 2022 }

\pubyear{2023}

\begin{document}
\label{firstpage}
\pagerange{\pageref{firstpage}--\pageref{lastpage}}
\maketitle

%Ása comment: From discussion: Ages of the stars, evolution sequence, spinstars, more elements.

\begin{abstract}

\noindent The origin of Carbon Enhanced Metal-Poor (CEMP-no) stars with low abundances of neutron-capture elements is still unclear. These stars are ubiquitous, found primarily in the Milky Way halo and ultra-faint dwarf galaxies (UFDs). To make a major step forward, we developed a data-calibrated model for Böotes~I that simultaneously includes all carbon sources: supernovae and asymptotic giant branch (AGB) stars both from first (Pop~III) stars, and subsequent normal star formation (Pop~II). We demonstrate that each of these sources leave a specific chemical signature in the gas, allowing us to identify the origin of present day CEMP-no stars through their location in the A(C)-[Fe/H] diagram. The CEMP stars with $\rm A(C)>6$ are predominantly enriched by AGB Pop~II stars. We identify a new class of {\it moderate CEMP-s} stars with A(C)$\sim$7 and $\rm 0<[Ba/Fe]<+1$, imprinted by winds from AGB stars. True Pop~III descendants are predicted to have A(C)<6 and a constant [C/Mg] with [Fe/H], in perfect agreement with observations in Böotes~I and the Milky Way halo. For the first time we now have a complete picture of the origins of CEMP-no stars which can and will be verified with future observations.

\end{abstract}


%%%%%%%%%%%%%%%%%%%%%%%%%%%%%%%%%%%%%%%%%%%%%%%%%%

%%%%%%%%%%%%%%%%% BODY OF PAPER %%%%%%%%%%%%%%%%%%

\section{Introduction}
Spectroscopic studies in the Galactic halo and in the Ultra-Faint Dwarf galaxies (UFDs) have revealed that a large fraction of very metal-poor stars, $\rm [Fe/H]<-2$, show an enhancement of carbon,  [C/Fe]\footnote{[X/Y] = $\rm \log10(m_{X}/m_{Y})$ - $\rm \log10(m_{X,\odot}/m_{Y,\odot})$, where $\rm m_{X}$ and $\rm m_{Y}$ are the abundances of elements X and Y and $\rm m_{X,\odot}$,  and $\rm m_{Y,\odot}$,  are the solar abundances of these elements \citep{Asplund+09}.}$> + 0.7$, i.e.~the so-called {\it Carbon Enhanced Metal-Poor} (CEMP) stars \citep{beers05,aoki14,Yong13,Norris13}. However, the origin of CEMP stars and their possible connection with the chemical elements produced by the first generation of stars (PopIII) is far from being fully understood. \\
CEMP stars are usually classified into two main classes: CEMP-$s$ ($r$) stars that show an enhancement in the slow (rapid) neutron-capture elements, and CEMP-no stars that do not show an excess in heavy elements ([Ba/Fe]< 0). 
The observed differences in the chemical signatures of these two classes seem to reflect the different origin of the carbon enhancement. In particular, CEMP- {\it s} are usually found in binary systems, therefore the observed enhancement of carbon and s-process elements is thought to be originated from mass transfer from a binary companion in the Asymptotic Giant Branch (AGB) phase (e.g. \citealt{starkenburg13,hansen15,Abate18}). On the contrary, CEMP-no stars are {\it not} preferentially found in binary systems and their peculiar abundance pattern is expected to be representative of the interstellar medium (ISM) from which they formed (e.g. Aguado et al. A\&A in press).

Observationally, \cite{bonifacio15} suggested an approach to  distinguish CEMP-no stars from  CEMP-{\it s} based on the absolute carbon abundance,\footnote{A(C) = $\log(\rm N_{C}/N_{H}) + 12 $ where $\rm N_{C}$ and $\rm N_{H}$ represent the number density fraction of carbon element and hydrogen, respectively.} A(C). The CEMP-{\it s} stars belong to the {\it{`high carbon band'}} with $\rm A(C) > 7.4$, while CEMP-no stars are characterized by  $\rm A(C) \leq 7.4$, i.e. they belong to the {\it{`low carbon band'}}. More recently, \cite{Yoon16} showed that CEMP-no stars can be further subdivided into two groups: Group II, which shows a correlation of A(C) with iron abundance $\rm [Fe/H]$, and Group~III, with nearly constant A(C) with [Fe/H].  

{\it But what is the origin of CEMP-no stars and the different nature of Group II \& III?} Many studies show that CEMP-no stars most likely formed from an ISM enriched by massive {\it{metal-free}} stars.

Furthermore the peculiar abundance patterns of CEMP-no stars belonging to Group III are consistent with a birth environment enriched by low-energy first supernovae (SNe), with mixing and fallback, the so-called {\it {faint supernovae}} \citep[e.g.][]{Iwamoto05, ishigaki14, SS15, Komiya_2020}, or fast rotating massive metal-free stars, the so-called {\it spinstars} 
\citep[e.g.][]{Meynet06, Liu21}. 
Lately a different view has been proposed in which excess carbon can be produced by {\it {normal}} PopII stars.
Some works showed that the carbon content of CEMP-no stars can be originated in a birth cloud enriched by AGB stars \citep[e.g.][]{Sharma19} or type~II SNe \citep{Jeon21}.  
Furthermore \cite{Chiaki+17} suggest that the difference between subgroups can be related to different gas coolants.
In conclusion, it is no yet clear what kind of stellar population was the main driver in the enrichment of CEMP-no stars, and whether different observed chemical abundance patterns are linked to different sources.

\noindent In this Letter we extend previous work \citep{Rossi+21} to investigate the origin of CEMP-no stars in UFDs, especially focusing on the best studied system, Boötes I.
 
\begin{figure}
\centering
\includegraphics[width=\columnwidth]{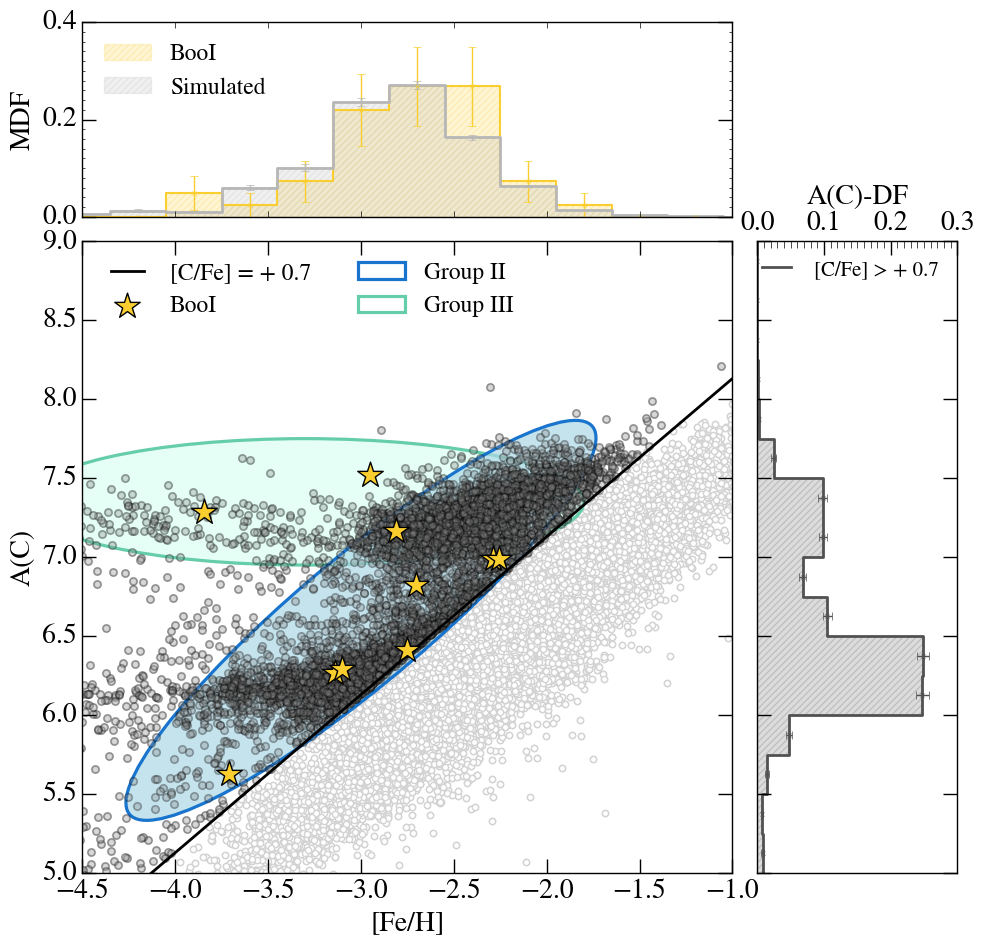}
\caption{The A(C)-[Fe/H] diagram of our simulated stellar populations (1000 realisations) of Boötes~I. The black line corresponds to [C/Fe] = +0.7.
Yellow star symbols identify the observed CEMP-no stars in Boötes~I, while the green and blue ellipses correspond to the CEMP-no groups by \citep{Yoon19}.
 The top plot shows the comparison between the observed (yellow; data from \citealt{norris10,lai11,gilmore13}) and simulated (grey) MDF. %of Boötes I. 
 The right marginal plot shows the simulated A(C)-DF of CEMP-no stars.}
\label{figure1}
\end{figure}  
\vspace{-0.6cm}
\section{Model Description} \label{Sec:Model}
\noindent The semi-analytical model, fully described in \cite{Rossi+21}, follows the star formation and the chemical enrichment history of a Boötes~I-like UFD from the epoch of its virialization until present day.
The initial conditions of the model have been set following the results of the cosmological model of \cite{SS09} and \cite{SS15}.
Following their results, we also assume an assembling history without major merger events, suitable for small galaxies such as Boötes~I \citep{Gallart+21}. 
The star formation rate is taken to be proportional to the mass of available gas, and the stellar mass formed is distributed according to an Initial Mass Function (IMF). Following the critical metallicity scenario, the transition from {\it{metal-free}} Pop~III to {\it{normal}} Pop~II/I stars is regulated by the metallicity of the ISM gas,\footnote{$\rm Z_{gas} = M^{ISM}_{Z}/M_{gas}$ where $\rm M^{ISM}_{Z}$ and $\rm M_{gas}$ are the mass of metals in the ISM and the gas mass, respectively.} $\rm Z_{gas}$.
Thus, Pop~III stars only form if $\rm Z_{gas} < Z_{cr} = 10^{-4.5} Z_{\odot}$ \citep{debennessuti17} according to a Larson IMF \citep{Lars98}:  $\phi(m_{\star}) = m^{-2.35}_{\star} exp(- m_{ch}/m_{\star})$ with $m_{ch}=10 \msun$ in the mass range $m_{\star}=[0.8-1000]\msun$ (see \cite{Rossi+21}). Instead Pop~II/I stars form if $\rm Z_{gas} > Z_{cr}$ according to Larson IMF with $m_{ch}=0.35 \msun$, in the mass range $m_{\star}=[0.1-100] \msun$. 

\noindent The model takes into account the incomplete sampling of the stellar IMF, essential to model the evolution of poorly star-forming UFDs \citep{Rossi+21}. Each time a star formation event occurs, a discrete number of stars is formed, with mass selected random from the assumed IMF. The chemical enrichment of the gas is then followed by taking into account the mass-dependent stellar evolutionary timescales of every single star. However, the model does not consider mass-transfer enrichment associated with binary systems. 

\noindent The model is data-calibrated: the free parameters have been fixed to reproduce the observed properties of Boötes~I, i.e. the total luminosity, the average stellar iron abundance, the Metallicity Distribution Function (MDF; see Fig.~\ref{figure1}), and the star formation history. Due to the stochastic nature of the IMF sampling, the chemical enrichment history of Boötes~I can slightly change between the different runs of the code. For this reason all the results presented here have been obtained by averaging over 1000 possible Boötes I evolutionary histories and by quantifying the scatter among them.

\subsection{New features}
\label{newfeatures}
To understand the origin of the carbon abundances measured in CEMP-no stars, we have expanded the model from \citet{Rossi+21} to track and to follow the evolution of all chemical elements from carbon to zinc, taking into account contributions from AGB stars and SNe. AGB stars in the mass range $[2-8]\msun$ enrich the ISM only through stellar winds (C, N, and O), and we adopt the stellar yields from \cite{VanDenHoek+97} for both Pop~III and Pop~II/I AGB stars. On the other hand, SNe can enrich the ISM with all the chemical elements from C to Zn. For Pop~II SNe in the mass range $[8-40]\msun$ we use the updated metal yields of \cite{Limongi+18} (model R, without rotation velocity) and an explosion energy, $\rm E_{SN} = 10^{51} erg$. We assume that Pop~III SNe in the mass range $[10-100]\msun$ can explode with different $\rm E_{SN}$ from $0.3 \times 10^{51}$ erg up to $10^{52}$~erg and different mixing efficiency $\eta = [0-0.25]$, and we adopt the corresponding yields provided by \cite{Heger+woosley10}. Every time that a Pop~III SNe forms we randomly assign to it an energy and a mixing efficiency. The energies allowed for Pop~III SNe are equiprobable and they determine the Pop~III SNe types:  {\emph{faint}} SNe with $ \rm E_{SN}=0.3-0.6 \times 10^{51}$~erg, \emph{core-collapse} SNe with $\rm E_{SN}=0.9-1.5 \times 10^{51}$~erg, \emph{high-energy} SNe  with $\rm E_{SN}=1.8-3.0 \times 10^{51}$~erg, and \emph{hypernovae} with  $\rm E_{SN}=5.0-10.0 \times 10^{51}$~erg.
For  massive Pop~III SNe in the mass range $[140-260]\msun$, the so-called {\it Pair Instability} SNe (PISN), we maintain the same assumptions of the previous work, i.e. an explosion energy proportional to the stellar mass, and we adopt the stellar yields of \cite{heger02}. 
    
\begin{figure*}
\includegraphics[width=\textwidth]{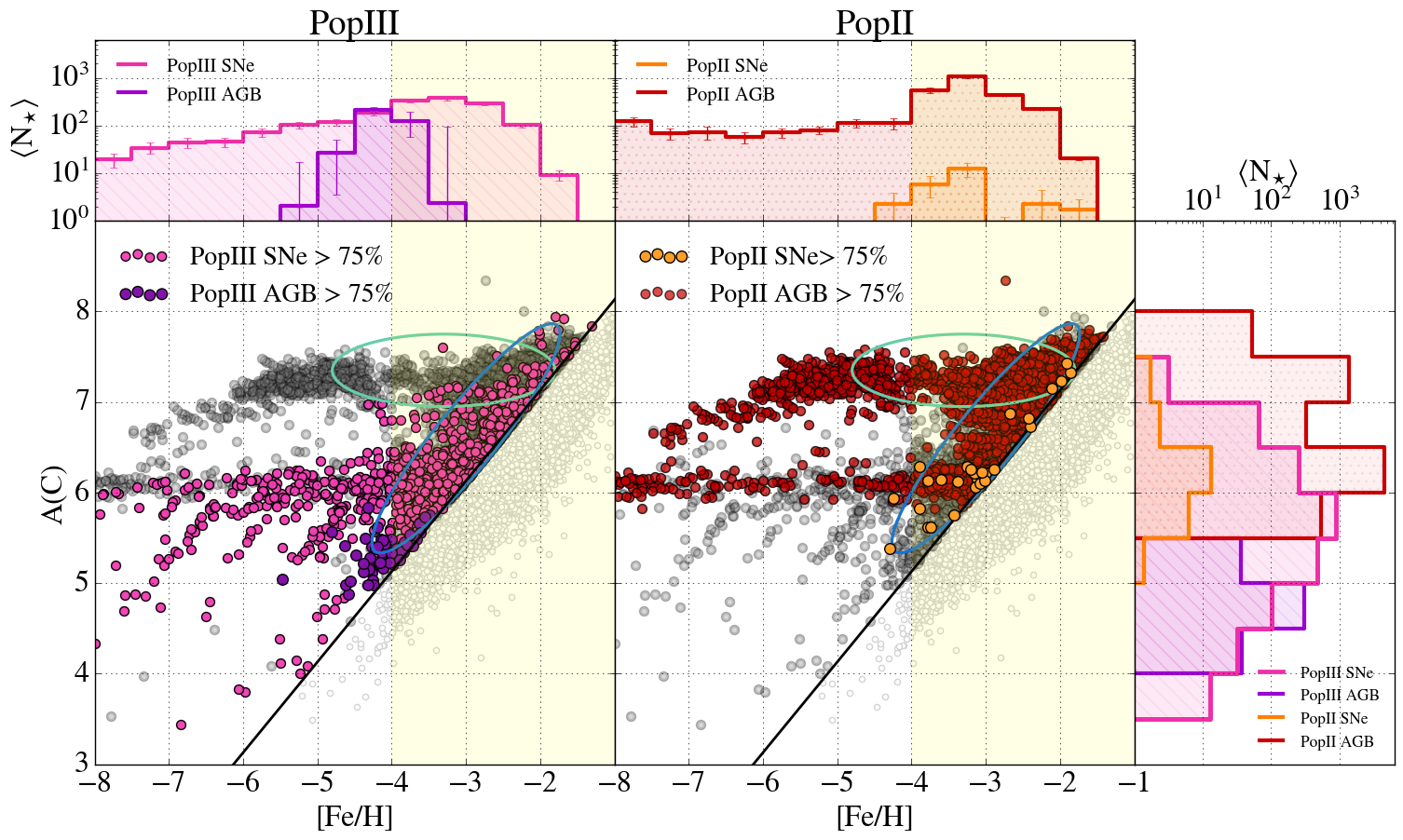}

\caption{The A(C) versus [Fe/H] diagram for different CEMP-no stellar populations. Colors highlight CEMP-no stars enriched to a level higher than $75\%$ by different stellar populations: in the columns we distinguish between Pop~III (left) and Pop~II stars (right) and in each panel we differentiate between SNe and AGB stars. In the marginal plots we show the number of stars averaged over the different runs, $\rm \langle N_{\star} \rangle $ as a function of [Fe/H] (top) and A(C) (right). The yellow shaded area represents the $\rm [Fe/H]$ range in which the currently available Boötes~I data are located. Note that all stellar populations have total metallicity $\rm Z_{\star} >\rm Z_{crit}$}.
\label{figure2} 
\end{figure*}

%\vspace{-0.65cm}
\section{Results}
\subsection{The carbon enrichment in Boötes~I}
In the top panel of Fig.\ref{figure1} we show a comparison between the observed MDF of Boötes~I and the simulated one, which are in good agreement, and the same is true for other observed properties (i.e. the total luminosity, the average stellar iron abundance and the star formation history). In fact, changing the properties of Pop~III stars (e.g. yields, IMF) has very limited effects on the MDF, since the general stellar population is dominantly enriched by normal Pop~II/I stars (see Sec.5 of \citealt{Rossi+21}). 

\noindent The central plot of Fig.\ref{figure1} shows the A(C) distribution with $\rm [Fe/H]$, for the simulated stellar populations of Boötes~I, where the CEMP-no stars ($\rm [C/Fe] > +0.7$) are highlighted (black symbols above solid line). 
In agreement with observational studies (e.g. \citealt{Yoon19}), we find the existence of two distinct groups (II \& III) which clearly show different behaviors in the A(C)-[Fe/H] diagram: Group~II exhibits a strong correlation with $\rm [Fe/H]$ while Group~III is constant with $\rm [Fe/H]$. The bimodal trend of A(C)
is even more evident if we look at the right marginal plot of Fig.\ref{figure1}, in which we show the A(C)-Distribution Function\footnote{A(C)-DF: the number of CEMP-no stars in bin of A(C) normalized with respect to the total number of CEMP-no stars} (A(C)-DF). As we can see the distribution is bimodal with two peaks at $\rm A(C) \sim 7.2$ and $ \rm A(C) \sim 6.2$ that roughly correspond to Group III and II, respectively.

\subsection{The different CEMP-no populations}
\noindent Following the evolution of single stars with different metallicity (Pop~III - Pop~II) and with different timescales (SNe - AGB) we can determine the main drivers of carbon enrichment in any given stellar population in Boötes~I.
Fig.\ref{figure2} shows the A(C)-[Fe/H] diagram for CEMP-no stars, where colors highlight what stellar population is the main driver of carbon enrichment: Pop~III or Pop~II stellar populations (left/right panels), SNe or AGB stars. The first thing to note is that CEMP-no stellar populations cover all the iron range $-8 < \rm [Fe/H] < -1$ regardless of which stellar population is the main driver of enrichment (Pop~III/Pop~II). It is also evident that two separate branches exist, both showing roughly constant $\rm A(C)$ values in the $\rm [Fe/H]$ range. The first branch, at $\rm A(C)
\sim 7 $, is populated by CEMP-no stars mainly enriched by Pop~II stars while, in the second branch, at $\rm A(C)\sim 6$, the enrichment is driven by Pop~III stars.

\begin{figure*}

\includegraphics[width=\textwidth]{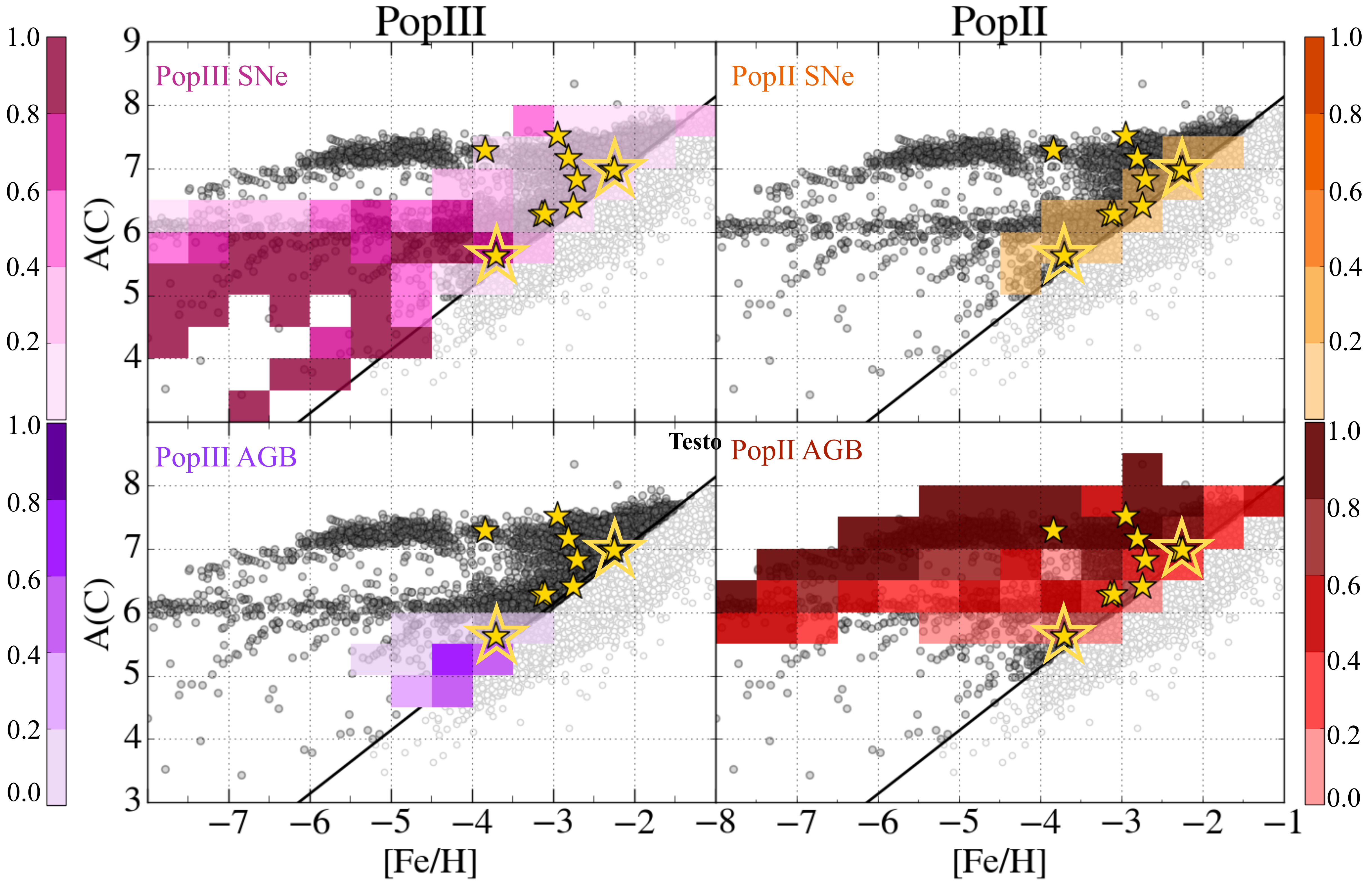}

\caption{The A(C) versus [Fe/H] diagram for CEMP-no stellar populations. The colored squares identify the probability to find,  at fixed [Fe/H] and A(C), CEMP-no stars predominantly enriched by Pop~III or by Pop~II stars (columns), while in the rows we distinguish between SNe and AGB stars. Yellow stars represent the Boötes~I data, while double marked points identify stars for which Mg abundance is available in the literature.}
\label{figure33}
\end{figure*}

\noindent In the range where data are available, $\rm[Fe/H]>-4$, the number of CEMP-no stars enriched by Pop~II stars, both Pop~II SNe (rapid timescale) and AGB stars (longer timescale), is $\sim 2$ times greater than those enriched by Pop~III stars (top marginal plot of Fig.\ref{figure2}), in good agreement with \citet{Sharma19}. It should be noted that almost all PopII-enriched CEMP stars have been polluted by AGB stars, while only a few ($\lesssim 1\%$) are PopII SNe enriched. The latter formed in birth cloud mainly enriched by {\it{normal}} PopII SNe but pre-polluted by PopIII SNe, allowing to reach [C/Fe] > +0.7.\\
On the other hand, at $\rm[Fe/H]\leq- 4$ the number of CEMP-no stars predominantly enriched by Pop~III stars is $\sim1.5$ times greater than those enriched by Pop~II stars. We find that these Pop~III star {\it{descendants}} formed out of an ISM enriched by low-energy SNe ($\leq 10^{51} \rm erg$). The branch-like behavior of PopIII enriched CEMP-no stars at $ \rm A(C)\sim6$ can be explained as follows. In our model for isolated UFDs the gas gradually accumulates and when the first Pop III stars form typically $\rm M_{gas} \gtrsim 10^3 \rm M_{\odot}$ \citep{Rossi+21}.  A single (or a few) low energy Pop III SNe produce a mass of carbon $\rm M_{\rm C} \sim 1 \: \rm M_{\odot} $, which is diluted in $\rm M_{gas} \sim 10^5 \rm M_{\odot} $ thus leading to $ \rm A(C)\sim6$.
Although the majority ($\sim 60\%$) of CEMP-no stars at $\rm[Fe/H]\leq-4$ have been enriched by the Pop~III SNe, $ \sim 40\%$ formed from an ISM mainly imprinted by Pop~II AGB stars. These 
%low-Fe stars predominantly imprinted by Pop~II AGB 
stars are rare, and form when most of the gas and metals have been ejected through previous SN explosions and thus AGB stars can substantially contribute to the ISM enrichment.
%As a consequence these stars are characterized by low [Fe/H] and high amount of carbon produced by AGB Pop~II stars. 
Fortunately, they can be easily distinguished from Pop~III enriched CEMP-no stars, thanks to their higher $\rm A(C)$ value. 
From the right marginal plot in Fig.\ref{figure2}, we clearly see that the distribution for CEMP-no stars enriched by Pop~II populations (SNe+AGB) covers the range $5.0 \leq \rm A(C)< 8.0$, with a peak at $ \rm A(C) \sim 6.5$, at which the enrichment is dominated by Pop~II AGB stars. On the other hand, the distribution for Pop~III (SNe+AGB) enriched CEMP-no stars covers the range $3.5 < \rm A(C) < 7.5$ with the peak at $ \rm A(C) \sim 6$.
Pop~III stars {\it{descendants}} are therefore characterized  by $ \rm A(C)\lesssim 6$ (and $\rm[Fe/H]\leq -3.5 $) and they are distinguishable from those mainly imprinted by Pop~II stellar populations identified by $ \rm A(C)>6$ for all $\rm [Fe/H]$. 

\noindent To unveil the origins of different CEMP-no stars, we derived the probability (at fixed [Fe/H] and A(C)) that CEMP-no stars have been imprinted by a given stellar population.
From Fig.\ref{figure33} it is evident that the probability to find Pop~III star {\it{descendants}} increases as the [Fe/H] decreases and it is maximum for $\rm A(C) < 6$ and $\rm [Fe/H]< -3.5$, where most of observational data for Boötes~I are not yet available, except for one star. For $\rm A(C)>6$ at all $\rm [Fe/H]$, the majority of CEMP-no stars are mainly enriched by Pop~II stars.
 Furthermore, we see that for $ \rm A(C) > 7$ and $\rm -4 <[Fe/H]< -1$, the probability to find CEMP-no stars mainly enriched by Pop~II AGB is $>80\%$ while by Pop~II SNe it is $ \lesssim 20\% $. Finally, by comparing our predictions with the available data in Boötes~I (Fig.\ref{figure33}) we can see that most of the observed CEMP-no stars are found at $\rm [Fe/H] > -4$ and $\rm A(C)> 6$, i.e. in the region where stars are more likely to have been imprinted by Pop~II stellar populations.
However, one star ($\rm [Fe/H] = -3.71 $, $\rm A(C)=5.62$; \citealt{norris10}) is found in a region where the probability that it has been Pop~III enriched is $\gtrsim 60\%$.
\vspace{-0.2cm}
\subsubsection{Moderate CEMP-s stars}
Since our model does not account for binary transfer (Sec.\ref{Sec:Model}), our results do not predict classical CEMP-s stars.   However, it is evident from Fig.\ref{figure2} that for $\rm6.5<A(C)<8$ there exists a class of CEMP stars forming from an ISM imprinted by Pop~II AGB stars, which represent about $\gtrsim 90\%$ of the Pop~II-enriched CEMP stars at $\rm [Fe/H]>-4$.  
These stars have different PopII AGB progenitors: stars with $\rm A(C) \sim 6.5 $ have been enriched on average by massive PopII AGB ($ m_{\star} \gtrsim 5 \rm M_{\odot}$),  while the ones with $\rm A(C) \gtrsim 7$ by low-massive PopII AGB ($m_{\star} < 5 \rm M_{\odot} $).The C-excess is {\it moderate} with respect to CEMP-s stars (A(C)$\sim$8), but it originates from the same source. Thus, we dub them {\it moderate CEMP-s stars}. These moderate CEMP-s stars are expected to have intermediate characteristics compared to CEMP-no ($\rm[Ba/Fe]<0$) and CEMP-s stars ($\rm [Ba/Fe]>+1$), because their chemical enrichment is not powered by mass transfer through a binary AGB companion. Our adopted AGB yields \citep{VanDenHoek+97} do not include s-process production, but by using observational constraints on the [Ba/C] ratio \citep{Masseron}  in combination with our predicted [C/Fe], we estimate these stars to have $\rm 0\lesssim[Ba/Fe]\lesssim+1$, filling the gap of unexplained CEMP stars. Our model is thus able to explain for the first time this neglected population of stars as {\it moderate CEMP-s} stars, formed from the products of Pop~II AGB stars. 

\subsection{Testing our predictions: [C/Mg]}

\begin{figure}
\centering
\includegraphics[width=\columnwidth]{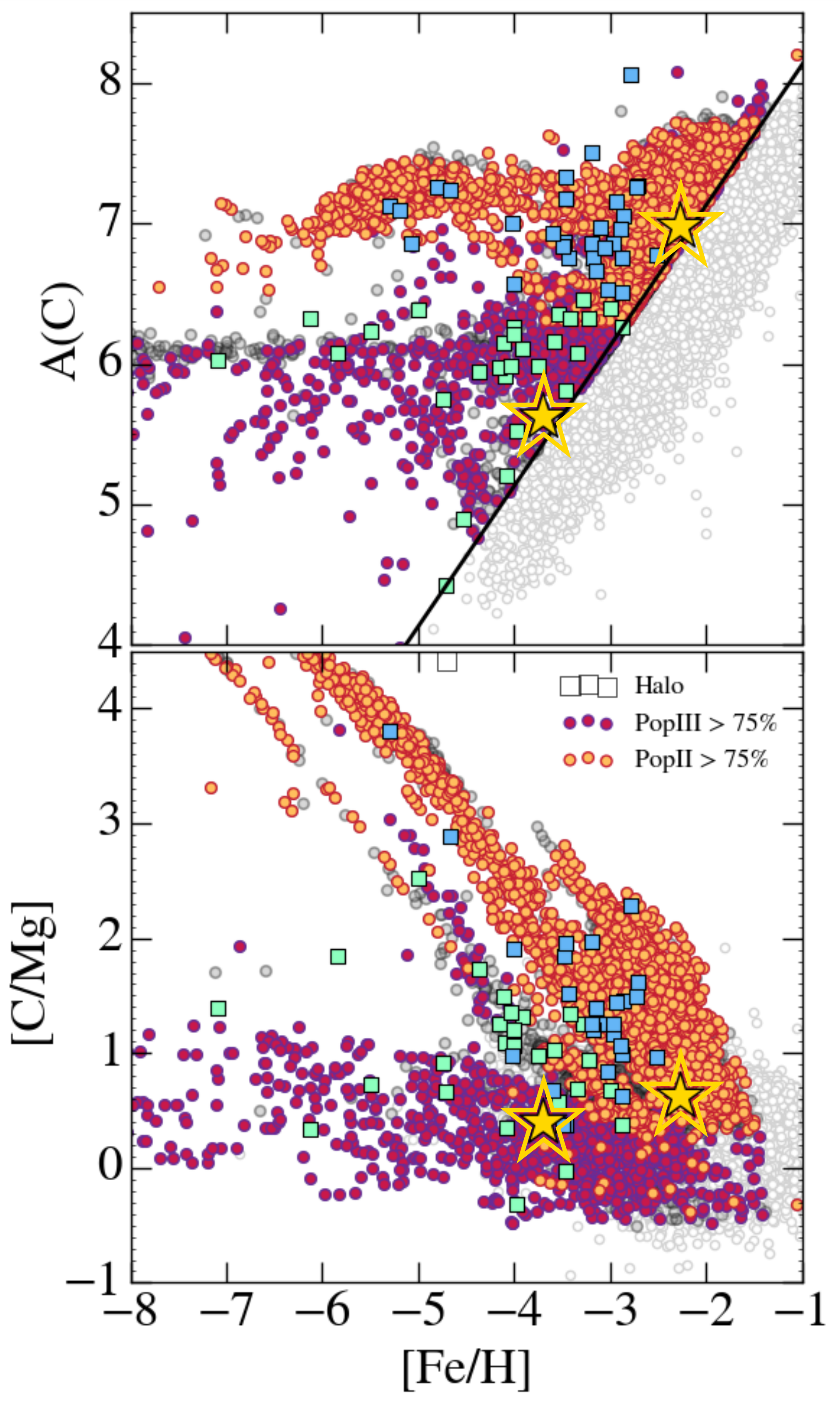}
\caption{The CEMP stars in Boötes~I, 
 enriched to a level higher than $75\%$ by PopIII (magenta) or by PopII (orange). Double marked yellow stars represent Boötes I stars for which there is available [Mg/Fe] \citep{norris10, chiti18}. Squares represent the Milky Way halo: $\rm A(C)>6$ (light blue), and $\rm A(C)<6$ (green), from the compilation of \citet{Yoon16}}. %\citep{Christlieb2004, Cayrel2004a, norris07, caffau11, yong+13, keller14, hansen15, Frebel15, bonifacio15, Li2015, Starkenburg18, bonifacio18, francois2018, aguado19, Ezzeddine2019} }
\label{figure4}
\end{figure}

The [C/Mg] ratio represents an additional key diagnostic on the origin of CEMP-no stars. In Fig.\ref{figure4} we identify two different [C/Mg] trends. The Pop~III enriched CEMP-no stars show an approximately constant [C/Mg] value over the entire [Fe/H] range. These stars formed in an environment predominantly enriched  by Pop~III SNe, which produce both C and Mg giving $\rm [C/Mg]\sim constant$, while their [Fe/H] depends on both the mass and explosion energy of Pop~III SNe (Rossi et al. in prep.). On the contrary, Pop~II enriched CEMP-no stars exhibit a strong correlation where [C/Mg] increases towards lower [Fe/H].  The position of these Pop~II enriched CEMP-no stars in [C/Mg]-[Fe/H] space therefore depends on the level of enrichment form PopII AGB stars: stars with $\rm [Fe/H]\leq -4 $ are `purely' Pop~II AGB imprinted ($\gtrsim 95\%$). They are therefore C-rich and Mg-poor giving high [C/Mg] values.  As [Fe/H] increases, CEMP-no stars form in environments mainly enriched by PopII AGB but with a significant contribution to the enrichment from PopII SNe ($\sim 25\%$).The latter mainly produce Mg, Fe giving a decrease of [C/Mg] ratio.

\noindent In Fig.~\ref{figure4} we compare the predicted [C/Mg] ratio for these two different stellar populations with the values measured in two CEMP-no stars in Boötes~I (double marked stars in Fig.\ref{figure33}, Fig.\ref{figure4}). These two stars separate very nicely into the two different stellar populations in our model: the star at the lowest value of [Fe/H] is consistent with an enrichment dominated by Pop~III stellar populations, while the other is compatible with Pop~II stellar populations enrichment. 
Finally, comparing our model predictions with Galactic halo data (squares in Fig.\ref{figure4}) we are able to probe the existence of the two branches in the A(C)-[Fe/H] diagram. This suggests that low-Fe ($\rm [Fe/H]<-4$) Pop~III imprinted CEMP-no stars could exist also in UDFs, unless they form in pre-enriched environments \citep[e.g.][]{SF12}. Furthermore, this comparison demonstrates that [C/Mg] is a key diagnostic tool to distinguish the origin of CEMP-no stars (see bottom panel of Fig.\ref{figure4}).

\section{Summary and discussion}
\label{summary}

For the first time we unveil the origin of CEMP-no stars by exploiting a self-consistent model \citep{Rossi+21}, data-calibrated on Boötes~I, that simultaneously accounts for the different enrichment sources of CEMP stars. We present a unique way of distinguishing between the primary sources of their enrichment (SNe or AGB stars) as well as the type of stellar populations (Pop~III or Pop~II) that polluted their natal clouds. Our model predicts where in the A(C)-[Fe/H] diagram present day stars should be located, depending on their main source of C-enrichment. The majority of CEMP stars at $ \rm A(C) \gtrsim 6$ are predominantly polluted Pop~II stars. In particular, we predict a population of CEMP stars, which get their C-enhancement from AGB stellar winds, without binary transfer. These {\it moderate CEMP-s stars} are expected to cover the range $\rm 0<[Ba/Fe]<+1$. Our model is thus able to explain these stars that fall in between the classical categories of CEMP-no and CEMP-s stars.

\noindent On the other hand, CEMP-no stars with $ \rm A(C) \lesssim 6$ are predominantly true Pop~III {\it{descendants}}. These Pop~III only-enriched CEMP-no stars, have roughly constant [C/Mg] with [Fe/H], in stark contrast with CEMP stars mainly enriched by Pop~II stars. Thus, [C/Mg] is a key diagnostic to distinguish between different populations. 
Furthermore, PopIII descendants are characterized by $\rm-0.5<[C/Mg]<+1$, in agreement with findings of \citet{Hartwig18}, despite different approaches.
Most of these Pop~III descendants lie at $\rm [Fe/H]<-4$, therefore representing a still hidden population in UFDs. However, observations of hyper iron-poor stars in the Milky Way strongly support our findings. This indicates a close connection of UFDs with the Milky Way halo at the lowest [Fe/H]. In the present and upcoming era of large spectroscopic surveys (e.g. 4DWARFS) and new instrumentation (e.g. ELT), we can expect find these rare populations, predicted by our model.
Only with dedicated models able to reproduce full abundance patterns (e.g. Rossi+23 in prep) will we be able to exploit this large influx of data to understand the properties of the first stars and the earliest galaxies.

\vspace{-0.65cm}
\section*{Acknowledgements}
We thank the anonymous referee for the useful and constructive comments.
This project has received funding from the European Research Council under the European Union’s Horizon 2020 research and innovation programme (grant agreement No 804240). S.S. and I.V. acknowledge support from the PRIN-MIUR2017, prot. n. 2017T4ARJ5. 
\vspace{-0.65cm}
\section*{Data AVAILABILITY}
All data are available on request from the corresponding author MR.
\vspace{-0.65cm}
%%%%%%%%%%%%%%%%%%%%%%%%%%%%%%%%%%%%%%%%%%%%%%%%%%



\bibliographystyle{mnras}
\bibliography{OriginOfCEMPno} % if your bibtex file is called example.bib

%%%%%%%%%%%%%%%%%%%%%%%%%%%%%%%%%%%%%%%%%%%%%%%%%%

%%%%%%%%%%%%%%%%% APPENDICES %%%%%%%%%%%%%%%%%%%%%

%%%%%%%%%%%%%%%%%%%%%%%%%%%%%%%%%%%%%%%%%%%%%%%%%%%%
% SAMPLING WITH MCH>1 AND FLAT IMF 
%%%%%%%%%%%%%%%%%%%%%%%%%%%%%%%%%%%%%%%%%%%%%%%%%%%%

%\appendix
%\section{}

% Don't change these lines
\bsp	% typesetting comment
\label{lastpage}

\end{document}